\documentclass[12pt]{article}
\usepackage{graphicx}
\usepackage{amsmath}
\usepackage{amsfonts}
\usepackage{amssymb}

\newcommand{\R}{\mathbb R}

\newcommand{\g}{\gamma}

\newcommand{\bg}{{\bar\gamma}}
\newcommand{\bK}{{\bar{K}}}
\newcommand{\bpi}{{\bar\pi}}
\newcommand{\bpsi}{{\bar\psi}}
\newcommand{\bn}{{\bar\nabla}}

\newcommand{\stg}{{\bf{g}}}

\newcommand{\stric}{\mbox{\bf{Ric}}}
\newcommand{\stscal}{\mbox{\bf{R}}}
\newcommand{\stG}{{\bf{G}}}
\newcommand{\stT}{{\bf{T}}}
\newcommand{\bgamma}{{\bar{\gamma}}}

\newcommand{\lr}{{\cal R}_{\gamma, \psi}}
\newcommand{\la}{{\cal A}_{\gamma, \tilde{K}, \pi}}
\newcommand{\lb}{{\cal B}_{\tau, \psi}}
\newcommand{\ulr}{{\underline{\cal R}}_{\gamma, \psi}}
\newcommand{\ula}{{\underline{\cal A}}_{\gamma, \tilde{K}, \pi}}
\newcommand{\ulb}{{\underline{\cal B}}_{\tau, \psi}}

\newcommand{\yam}{{\mathcal Y}_{\psi}([\gamma])}

\newcommand{\vp}{\varphi}

\newcommand{\di}{\hbox{div}}
\newcommand{\divg}{{\mbox{\rm div}_{\stg}}}

\newcommand{\tr}{{\mbox{\rm tr\,}}}

\newcounter{shownewstuffflag}

\newcommand{\startnewstuff}{\ifnum\value{shownewstuffflag}>0\color{blue}\fi}
\newcommand{\finishnewstuff}{\ifnum\value{shownewstuffflag}>0\color{black}\fi}

\newcounter{oldeq}

\newcounter{mnotecount}[section]

\newcommand{\rmnote}[1]{}

\def\beq{\begin{equation}}
\def\eeq{\end{equation}}

\newtheorem{theorem}{Theorem}[section]

\newtheorem{corollary}[theorem]{Corollary}
\newtheorem{remark}[theorem]{Remark}
\newenvironment{proof}[1][Proof]{\textsc{#1.} }{\ \rule{0.5em}{0.5em}}
\numberwithin{equation}{section}

\begin{document}

\title{Applications of theorems of Jean  Leray  to the Einstein-scalar field equations}
\author{Yvonne Choquet-Bruhat 
\\ University of Paris \and
James Isenberg\thanks{Partially supported by the NSF under Grant
 PHY-0354659 }
\\ University of Oregon  \and
Daniel Pollack\thanks{Partially supported by the NSF under Grant DMS-0305048}
\\ University of Washington}

\date{November 1, 2006}

\maketitle

\centerline{\textit{Dedicated to the memory of Jean Leray}}

\begin{abstract} 
The Einstein-scalar field theory can be used to model gravitational physics with scalar field matter sources. We discuss the initial value formulation of this field theory, and show that the ideas of Leray can be used to show that the Einstein-scalar field 
system of partial differential equations is well-posed as an evolutionary system. We also show that one can generate solutions of the Einstein-scalar field constraint equations using conformal methods.

\end{abstract}

\section{Introduction}

In Newtonian theory, one models gravitational physics by studying a linear elliptic Poisson equation for the Newtonian potential on a fixed absolute background space and time, with the motion of material bodies  governed by the Newtonian force equation on this fixed background. By contrast,  in general relativity the gravitational field is modeled using Lorentzian spacetimes whose curvature reflects the material and field content of the spacetime. Mathematically, a Lorentzian spacetime is a pair $(M^{n+1}, \stg)$ where $M^{n+1}$ is a smooth manifold of dimension $n+1$ (in everyday physics $n=3$, but higher dimensions are sometimes considered for modelling electromagnetism and other interactions with gravity), and $\stg$ is a pseudo-Riemannian metric of signature $(-,+,\ldots,+)$. The metric distinguishes timelike directions $\stg(X,X)<0$ for a tangent vector $X$ (along the possible path for a massive physical object), null directions $\stg(X,X)=0$ (along the possible path for a massless physical particle), and spacelike directions $\stg(X,X)<0$. The physical time between a pair of events, as would be marked by a proper clock, corresponds to the $\stg$-length of the timelike trajectory followed by that clock. The path followed by a test particle which is free of non gravitational forces  corresponds to a timelike geodesic in the spacetime. 

It is well known that theoretical studies of general relativity have predicted such strange and interesting phenomena as the expansion of the universe, black holes, gravitational lenses, and gravitational waves. These phenomena, all of which have now been confirmed either by direct or indirect observation, were originally discovered via studies of solutions of the Einstein field equations. It is also well known, at least among mathematicians, that Einstein's equations present a number of very challenging mathematical problems, such as the cosmic censorship conjectures and the question of the nonlinear stability of black holes. 

In this brief review, we show that in two important ways, the ideas of Leray have played an important role in the study of Einstein's equations. In previous studies  these ideas were applied in the absence of a 
scalar field. Here, since scalar fields are now viewed as possibly important for understanding the apparent acceleration of the expansion of the universe, and since including them results in some additional interesting features in the analysis, we work with the Einstein-scalar field system. We introduce this system in section 2 and discuss its Cauchy formulation in section 3. We then show in section 4 how Leray's ideas play a role in understanding the evolutionary aspect of the Einstein-scalar field system. In section 5 we prove the existence
of  solutions to the Einstein-scalar constraint equations via the conformal methods, where seminal ideas of 
Leray regarding solutions of nonlinear elliptic equations have played a crucial role.

We end this section with a note about notation.  Wherever there may arise a possible confusion, we use bold faced
symbols for spacetime variables and tensors.  For initial data sets, we reserve the notation
of over-barred symbols for physical variables, which satisfy the relevant constraint equations, and may or may 
not be ``time-dependent" depending on the context.  Unadorned
symbols are used for free conformal data as described in section 5.  Finally the occasional appearance of 
a   ``tilded"-symbol (as in (\ref{detK}) below) appears when we need to introduce an intermediate quantity 
which is neither a free variable, nor a physical one.

\medskip
\noindent
{\bf Acknowledgement:}  The authors would like to thank the Isaac Newton Institute of Mathematical 
Sciences in Cambridge, England for providing an excellent research environment  during the program 
on Global Problems in Mathematical Relativity during Autumn, 2005, and again in October 2006, 
where some of  this research was carried out.

\section{\textbf{Einstein - scalar field equations}.}

For general source fields, the Einstein gravitational field equations take the tensorial form 
\begin{equation} 
\label{Eineq}
\stG(\stg)=  \stT(\Phi, \stg),
\end{equation}
where $\stG(\stg)$ is the Einstein tensor, which is a second order differential operator on the metric defined by 
$\stG(\stg) :=\stric(\stg) -\frac{1}{2}\stscal\,\stg$ with $\stric(\stg)$ denoting the Ricci tensor of $\stg$ and 
$\stscal$ denoting the scalar curvature of $\stg$, and where $\stT(\Phi,\stg)$ is the stress-energy or energy-momentum tensor\footnote{Note that we have chosen units so that $8\pi$ times the Newtonian gravitational constant is set equal to one.}, a specified functional of the matter source fields $\Phi$ and the metric. The specific form that the stress-energy tensor takes depends upon the matter source fields presumed to be present in the physical system being modeled. Here, we presume that ${\Phi}$ is a scalar field, which we label ${\Psi}$, with potential function 
$V(\Psi)$, and we set 
\begin{equation}
\label{Scals-e} 
\stT= \partial \Psi \otimes \partial \Psi - \left[ \frac{1}{2} |\partial \Psi|_\stg^2 + V(\Psi)\right] \stg. 
\end{equation} 

In addition to the equation (\ref{Eineq}), the Einstein-scalar field theory includes a field equation for $\Psi$, which reads
\begin{equation}
\label{Psieq}
\nabla^{\alpha}\partial_{\alpha}\Psi=\frac{dV}{d\Psi}, 
\end{equation}
where $\nabla$ denotes the covariant derivative compatible with $\stg$. While this extra equation may simply be added to the theory by hand, it also follows directly as a necessary consequence of equation (\ref{Eineq}) together with the geometric properties of the Einstein tensor: One readily verifies that the Bianchi identities for the curvature imply that the Einstein tensor satisfies the identity 
\begin{equation} 
\divg \stG(\stg) =0
\end{equation}
where $\divg$ denotes the divergence operator for the metric $\stg$.
This condition together with (\ref{Eineq}) implies the conservation law
\begin{equation}
\label{Conseq} 
\divg \stT = 0. 
\end{equation}
We readily verify that (\ref{Conseq}) applied to the scalar field stress-energy tensor (\ref{Scals-e}) results in the field equation (\ref{Psieq}).

How does one choose the scalar field potential function $V(\Psi)$? While there are many possibilities, we note that 
$V(\Psi)=\frac{m}{2}\Psi^2$ corresponds to  the massive Klein-Gordon field, while setting $V(\Psi)=\Lambda$ for a non zero constant $\Lambda$ and requiring that $\Psi=0$ produces the vacuum Einstein theory with non zero cosmological constant 
$\Lambda$.

\section{The Cauchy problem: constraints and evolution.}

The Einstein-scalar field system of partial differential equations on a $3+1$ spacetime consists of eleven equations (\ref{Eineq}) and (\ref{Psieq}) for the eleven field variables $\stg_{\mu \nu}$ and $\Psi$ (for $n+1$ dimensions, there are $\frac12 (n+1)(n+2) +1$ equations for the same number of field variables). One of its most characteristic features, however, is that it is both an under-determined and an over-determined system, in the sense that if one formulates the Einstein-scalar-field system as a Cauchy problem, there are constraint equations which must be satisfied by any candidate set of initial data, and as the data evolves there are certain of the field variables whose evolution is entirely at one's discretion. Both of these features reflect the spacetime covariance of the theory (i.e., the theory has the spacetime diffeomorphism group as its gauge group).

To see these features explicitly, we now sketch out an $n+1$ decomposition of the Einstein-scalar field variables and equations. Given a spacetime $(M^{n+1}, \stg)$, we start by choosing an $n+1$ foliation of the spacetime manifold $F_t: \Sigma^n \rightarrow M^{n+1}$ $(t \in \R)$, for which each of the leaves $F_t(\Sigma^n)$ of the foliation is presumed spacelike. We also choose a threading of the spacetime by a congruence of timelike observer paths $T_{p}: \R\rightarrow M^{n+1}$ $(p\in \Sigma^n)$. The choice of a foliation and a threading, together with a choice of coordinate patches for $\Sigma^n$, automatically determines local coordinates 
$(x^0=t, x^1,\ldots, x^n)$ and local coordinate bases $(\frac{\partial}{\partial t}, \frac{\partial}{\partial {x^1}} \ldots, \frac{\partial}{\partial {x^n}})$ covering $M^{n+1}$. We may then, without loss of generality, write the metric (locally) in the form
\[
\stg=-N^2 \theta^t \otimes \theta^t + \bgamma_{ij} \theta ^i \otimes \theta^j
\]
where $(\theta^t=dt, \theta^j = dx^j + \beta ^j dt)$ is the one-form basis dual to the surface-compatible tangent vector basis $(e_{\perp}= \frac{1}{N}(\frac{\partial}{\partial t}-\beta^j \frac{\partial}{\partial x^j}), \frac{\partial}{\partial {x^1}} \ldots, \frac{\partial}{\partial {x^n}})$ with $e_{\perp}$ the normal vector field to $F_t(\Sigma^n)$. Here $N$ is the positive definite ``lapse function", $\beta^j$ are the components of the spacelike ``shift vector", and $\bgamma_{ij} $ are the components of the spatial metric tensor. We note that for each choice of $t$, $\bgamma(t)= \bgamma_{ij}(t) dx^i \otimes dx^j$ is the induced Riemannian metric on the leaf $F_t(\Sigma^n)$. We use the notation $\bK(t) = \bK_{ij}(t) dx^i \otimes dx^j$  to denote the second fundamental form defined by the foliation.

For the scalar field $\Psi$, there is no need to do any space + time decomposition. However, we shall use the notation $\bpsi$ to denote the restriction of $\Psi$ to one of the leaves of the chosen foliation, and we shall use the definition  $\bpi:=\frac{1}{N}(\frac{\partial}{\partial t }\bpsi-\beta^m \frac{\partial}{\partial x^m} \bpsi)$ for convenience in working with the time derivative of $\Psi.$

If we now use the usual $n+1$ decomposition to express the spacetime curvature in terms of the time dependent spatially covariant quantities $\bg, \bK, N, \beta$ and their various derivatives and (spatial) curvature, we find that the Einstein-scalar field equations (\ref{Eineq}) and (\ref{Psieq}) split into two types: constraint equations which require any choice of initial data to satisfy certain identities, and evolution equations which describe how the spatial fields evolve from one leaf of the foliation to the others. Explicitly, we have the following:

\medskip
\noindent\textbf{Constraint Equations}\\
From the $\stG_{\perp \perp}$ equation derived from (\ref{Eineq}), we obtain the {\it Hamiltonian constraint}
\begin{equation}
\label{hce}
2N^{-2}\stG_{\perp \perp}\equiv R_{\bgamma} -|\bK|_{\bgamma}^2 + (tr \bK)^2= \bpi^2 +|\bn \bpsi |_{\bgamma}^2 +2 V(\bpsi).
\end{equation}
From the $\stG_{\perp j}$ equations derived from (\ref{Eineq}), we obtain the {\it momentum constraint}
\begin{equation}
\label{mce}
-N^{-1}\stG_{\perp j}\equiv\bn_m\bK^m_j -\partial_j tr\bK =\bpi \partial_j \bpsi.
\end{equation}
Note that these equations constrain the choice of the data $(\bgamma, \bK, \bpsi, \bpi)$; they do not involve the lapse and shift.
We refer the interested reader to \cite{BI04} for a survey on the constraint equations.

\medskip
\noindent\textbf{Evolution Equations}\\
From the $\stG_{ij}$ equations derived from (\ref{Eineq}),  we obtain
\begin{eqnarray}
\frac{\partial}{\partial t }\bK_{ij} &=&  N (R_{ij} -2\bK_{im}\bK^m_j +\tr \bK \bK_{ij} -\partial_i \bpsi \partial_j \bpsi +\frac{1}{n-1} \bgamma_{ij} V(\bpsi))\nonumber \\
&&-\frac{1}{N} \bn_i \partial_j N +{\cal L}_\beta \bK_{ij},
\label{KEv}
\end{eqnarray}
where here and above $\bn$ is the covariant derivative associated to $\bg$, $R_{ij}$ are the components of the spatial Ricci tensor calculated from $\bg$, and ${\cal L}$ denotes the Lie derivative operator. This is an evolution equation for $\bK$. We obtain an evolution equation for $\bpi$ from the spacetime field equation (\ref{Psieq}) for $\bpsi$ 
\begin{equation} 
\frac{\partial}{\partial t }\bpi = N( \Delta_{\bgamma} \bpsi +  \tr \bK \bpi -\frac{dV}{d\bpsi})  + \bgamma^{mn} \partial_m N \partial_n \bpsi + {\cal L}_\beta  \bpi,
\end{equation}
where $\Delta_{\bgamma}$ is the Laplace-Beltrami operator for the metric $\bgamma$. 

We have evolution equations for $\bK$ and $\bpi$. What about $\bgamma, N, \beta$ and $\bpsi$? The evolution equation for $\bpsi$ comes from the definition for $\bpi$: 
\begin{equation}
\frac{\partial}{\partial t } \bpsi = N\bpi +{\cal L}_\beta \bpsi. 
\end{equation}
The evolution equation for $\bgamma$ comes from the definition of the second fundamental form:
\begin{equation} 
\frac{\partial}{\partial t} \bgamma_{ij}= -2N \bK_{ij} +{\cal L}_\beta \bgamma_{ij}.
\label{gammaEv}
\end{equation}
For the other field variables, $N$ and $\beta$, there are no evolution equations. This freedom to choose $N$ and $\beta$ and their evolution any way one wishes reflects the gauge invariance of the field equations under the action of the diffeomorphism group.

To summarize, the Cauchy formulation of the Einstein-scalar field equations asks that one choose the initial data set $(\bgamma, \bK, \bpsi, \bpi )$ subject to the constraint equations (\ref{hce})-(\ref{mce}). One then chooses $N$ and $\beta$ freely in time, to fix the gauge, and finally one proceeds to evolve $(\bgamma, \bK, \bpsi, \bpi )$ via the evolution equations just listed. Note that if one chooses the constraints to hold initially, in any accurate evolution  they must remain satisfied for all time.  

\section{Evolution system.}

Does an initial data set satisfying the constraint equations (\ref{hce})-(\ref{mce}) always generate (via the evolution equations (\ref{KEv})-(\ref{gammaEv})) a spacetime solution of the Einstein-scalar field equations (\ref{Eineq})-(\ref{Psieq})? To show this, one needs to prove that the system is well-posed in some appropriate sense. Here we discuss the well-posedness results of Leray-Ohya and of Leray, as applied to the Einstein-scalar field system.  Note that our focus here is solely on the evolution system.
As in the vacuum case  for the wave gauge \cite{FB52}, one must use the Bianchi identity to show that the constraints are
preserved in the evolution, thus yielding a local existence result for the full field equations.  We refer the reader to 
\cite{AACBY} where this issue (again for vacuum solutions) is addressed in a similar setting to the one considered here. 

To check if a given system is hyperbolic in the Leray-Ohya sense, one seeks to diagonalize the matrix of the principal parts (highest derivatives) of the system. If this diagonalization can be done, the system is called ``quasi-diagonal'',  and is a Leray-Ohya hyperbolic system if in addition each operator has a characteristic cone which contains the metric cone. As a consequence, it can be shown that it is well-posed in Gevrey classes of functions.  These are spaces of $C^{\infty}$ functions whose successive derivatives satisfy certain inequalities which are generally too weak to imply the convergence of the corresponding
Taylor series. Well-posedness means that a set of initial data in such classes will generate a spacetime solution with evolving data in the same classes. It also implies causal propagation, with the domain of dependence of the data determined by the causal cones of the spacetime metric, as well as continuous dependence of the evolved solution on the choice of data. Note that the principal parts of a Leray-Ohya system may generally have multiple characteristics. Also note that in verifying the criteria for a Leray-Ohya system, one need not have the  same 
order for each of the various evolution equations which make up the system. 

If in fact the operators in the principal part of a Leray-Ohya system do not have multiple characteristics --
i.e., if there exists a cone in the cotangent plane
such that each straight line passing through a point in its interior cuts the
characteristic cone in $N$ distinct points if the operator is of order $N$, then the system is called Leray  hyperbolic. 
It can be shown that such a system is well-posed in Sobolev spaces as well in Gevrey class spaces. 

For a more detailed discussion of Leray-Ohya and Leray hyperbolicity, see \cite{Ler, LO, CB06}.

We now apply these ideas to the Einstein-scalar field system. In doing so, we work with the Einstein-scalar field
system in mixed first order (for $\bg$ and $\bK$) and second order (for $\bpsi$), to mesh with the extant treatments. 

\subsection{Leray - Ohya hyperbolic system for $\bg,\bK$ and $\bpsi.$}

We set $\bar{\partial}_{0}:=\frac{\partial}{\partial t}-\mathcal{L}_\beta$ and
we consider the system

\[
\bar{\partial}_{00}^{2}R_{ij}-\bar{\partial}_{0}\bar{\nabla}_{i}R_{j0}
-\bar{\partial}_{0}\bar{\nabla}_{j}R_{i0}+\bar{\nabla}_{i}\bar{\nabla}
_{j}^{\text{ }}R_{00}=F_{ij}
\]
which is derived  by taking linear combinations of equations from the
Einstein-scalar field system (see \cite{AACBY} for a similar derivation in the vacuum case).
If  $\bar{\partial}_{0}\bg_{ij}$ is replaced by its value in terms of $\bK$,
\begin{equation}
\bar{\partial}_{0}\bg_{ij}=-2N\bK_{ij},
\label{41}
\end{equation}
this system reads as a third order system for $\bK$ of the form
\begin{equation}
\bar{\partial}_{0}{}(-N^{-2}\bar{\partial}_{0}^{2}+\bar{\nabla}^{h}\bar
{\nabla}_{h})\bK_{ij}=f_{ij}\,(2\text{ in }\bg, 2\text{ in }\bK, 3\text{ in }N, 3\text{ in }\bpsi) +\tilde{f}_{ij},
\label{42}
\end{equation}
with
\[
\tilde{f}_{ij}:=N\bar{\nabla}_{i}\partial_{j}(N^{-2}\bar{\partial}_{0}
^{2}-\bar{\nabla}^{h}\partial_{h})N. 
\]
The numbers appearing above with $f_{ij}$ (and below with $h$) denote the order of the
highest derivatives of the unknowns which  occur in that term. The additional
term $N\bar{\nabla}_{i}\partial_{j}(N^{-2}\bar{\partial}_{0}^{2}-\bar{\nabla}^{h}\partial_{h})N$ 
is clearly third order in $\bg$ (and fourth order in $N$).

The wave equation for $\bpsi$ reads, in terms of $\bg$, $\bK$ and the
presumably specified variables $N$ and $\beta$, as follows:
\[
-N^{-1}\partial_{0}(N^{-1}\partial_{0}\bpsi)+N^{-1}\bg^{ij}\bar{\nabla}
_{i}(N\partial_{j}\bpsi)+\bK_{i}^{i}N^{-1}\partial_{0}\bpsi=\frac{dV}{d\bpsi}.
\]
Applying the operator $\partial_{0}$ and using (\ref{41}) gives an equation of the
form 
\begin{equation}
\partial_{0}(-N^{-2}\partial_{0}^{2}+\bar{\nabla}^{h}\partial_{h})\bpsi=h\,(1
\text{ in }\bg, 1\text{ in }\bK, \text{ }2\text{ in }N, \text{ }2\text{ in }
\bpsi).   \label{43}
\end{equation}

The principal matrix of a system of partial differential equations $E_{B}(u_A)=0$, with unknowns
$u_{A}$, is obtained by assigning to each unknown an integer $m(u_{A})$ and to
each equation an integer $n(E_{B})$ such that the highest derivatives of
$u_{A}$ appearing in $E_{B}$ are at most of order $m(u_{A})-n(E_{B})$. The
principal part relative to $u_{A}$ in the equation $E_{B}=0$  then consists of the
terms of order $m(u_{A})-n(E_{B})$ in $u_{A}.$ It is zero if there are only
terms of smaller order. These integers $m(u_{A})$ and $n(E_{B})$ are collectively called the Leray-Volevic indices for the system.

For the system of PDEs  (\ref{41}), (\ref{42}) and (\ref{43})  with the 
field variables   $\bg(t)$, $\bK(t)$ and $\bpsi(t)$, 
we choose the Leray-Volevic indices
\begin{equation}
m(\bg)=3,\quad m(K)=3,\quad m(\Psi)=4,
\end{equation}
\begin{equation}
n(\ref{41})=2,\quad n(\ref{42})=0,\quad n(\ref{43})=1.
\end{equation}
The principal matrix is then a triangular matrix, with the elements in the diagonal
being  the  derivative $\bar{\partial}_{0}$ along the normal $e_{\perp}$ to the space
sections and the product of this operator by the wave operator in the
spacetime matrix. These operators are causal and hyperbolic. Both are such
that their characteristic cones at a point contain the metric null cone. 
However the non-diagonal form of the principal matrix does not permit one to
conclude that it is Leray  hyperbolic. 

If we replace $\bK$ in (\ref{42}) by its
value in terms of $\bg$ using (\ref{41}), and give to $\bg$ the index 4, we obtain
for (\ref{42}) and (\ref{43}) a diagonal system with principal operators the wave operator
and the operator
\[
\bar{\partial}_{0}^{2}{}(-N^{-2}\bar{\partial}_{0}^{2}+\bar{\nabla}^{h}
\bar{\nabla}_{h}).
\]
This operator has a double characteristic, the spacelike hyperplane, and the
system is only Leray-Ohya hyperbolic,  relative to the Gevrey class of
index 2. The Cauchy problem for this system is well posed in this class; the
domain of dependence of the solution is determined by the light cone of the
spacetime metric.

\subsection{Leray hyperbolic system for $\bg$, $\bK$ and $\bpsi$ with a lapse condition.}

The system for $\bg$, $\bK$ and  $\bpsi$ can be put into Leray hyperbolic form if we impose 
 a condition on the lapse function $N$ which makes it a quasi-diagonal system
for $\bg$, $\bK$,  $\bpsi$  and now also $N.$

To remove the term $\tilde{f}_{ij}$, which introduces non diagonal elements
into  the principal matrix, we require that $N$  satisfy a wave equation with source term, with that
source term being  an arbitrarily specified  function $F$
\[
\label{NEq}
-N^{-2}\partial_{00}^{2}N+\bar{\nabla}^{i}\partial_{i}N=F,
\]
and we then insert this equation into (\ref{42}). We now consider the system consisting 
of (\ref{41})-(\ref{43}) with this change,  together with the equation
\begin{equation}
\bar{\partial}_{0}((-N^{-2}\bar{\partial}_{0}^{2}+\bar{\nabla}^{i}\partial
_{i})N)=\bar{\partial}_{0}F,
\label{46}
\end{equation}
obtained by taking the $\bar \partial_0$ derivative of (\ref{NEq}), and using (\ref{41}).
We choose for   $N$ and (\ref{46}) the Leray-Volevic indices
\begin{equation}
m(N)=4,\quad n(\ref{46})=1.
\end{equation}
The system is now quasi-diagonal with hyperbolic diagonal elements given by
$\bar{\partial}_{0}(-N^{-2}\bar{\partial}_{0}^{2}+\bar{\nabla}^{i}\bar{\nabla
}_{i}) $ and $\bar{\partial}_{0}$. This leads to the following result.

\begin{theorem}
The system (\ref{41}), (\ref{42}), (\ref{43}), and (\ref{46}) is a Leray causal hyperbolic system for 
$\bg, \bK, \bpsi$ and $N$.
\end{theorem}

\section{The constraints}
\label{constraints}

The Einstein-scalar field constraints consist of the $n+1$ equations (\ref{hce})-(\ref{mce}), to be satisfied by the 
initial data $(\bg, \bK,\bpsi, \bpi)$ on an $n$-dimensional manifold $\Sigma$.
Locally, due to the symmetry of these tensors, this initial data can be regarded 
as a set of  $n(n+1)+ 2$ functions, which makes the underdetermined 
nature of the constraint equations apparent.
We recall that $(\bar{\g}, \bar{K}, \bar{\psi},  \bar{\pi})$ denotes a set of initial data which satisfies  the constraint equations; we use the same quantities without the over bars to denote functions which we choose freely in order to construct the data $(\bar{\g}, \bar{K}, \bar{\psi},  \bar{\pi})$. 
Also for convenience here, we restrict our considerations to the  so called
cosmological case, for which  $\Sigma$ is a compact manifold (see \cite{CBIP} for a 
treatment of the asymptotically flat case). Even in vacuum there are
infinitely many solutions of the constraints, depending on arbitrary transverse-traceless
(divergence and trace free)
tensors, which can be interpreted as ``radiation data''. 

\subsection{The conformally formulated constraints}

The conformal method  involves decomposing the data  $(\bar{\g}, \bar{K}, \bar{\psi},  \bar{\pi})$ into certain parts which are chosen freely, and other parts which are determined by solving equations which we derive from the constraint equations. We consider the case $n\geq3$.
The most basic piece of the freely chosen data is a choice of a Riemannian
metric $\gamma$, or rather the conformal class of metrics represented by $\gamma$. 
The physical metric $\bar{\g}$ is required  to be conformally related to $\gamma$. 
One sets
\[
\bar{\g}\equiv\vp^{\frac{4}{n-2}}\gamma
\]
for a positive function $\varphi$ on $\Sigma$.
The following identity  then holds between the scalar curvatures of $\bar{\g}$ and
$\gamma$
\begin{equation}
\label{cscal}
R(\bar{\gamma})=-\varphi^{-\frac{n+2}{n-2}}\left({\frac{4(n-1)}{n-2}}\Delta_{\gamma}\varphi -  R(\gamma)\varphi\right).
\end{equation}
On the other hand, the divergences of traceless contravariant symmetric
2-tensors are related by the identity
\begin{equation}
\label{cdiv}
\di_{\bar{\g}}\bar{P}=\varphi^{-\frac{2(n+2)}{n-2}}\di_{\g}P
\end{equation}
if
\[
\bar{P}=\varphi^{-\frac{2(n+2)}{n-2}}P.
\]
Using (\ref{cdiv}) (applied to the traceless part of $\bar{K}$) together with (\ref{cscal}), the Einstein-scalar field constraints may be conformally
reformulated as follows.

\subsubsection{The Momentum constraint}
The momentum constraint may be expressed with respect to the background metric $\vp$ by
\begin{equation}
\di_{\gamma} \tilde{K}=\frac{n-1}{n}\varphi^{\frac{2n}{n-2}}\nabla\tau
+\varphi^{\frac{2(n+2)}{n-2}}\bar{J},
\label{momeqn}
\end{equation}
where $\tau=\tr_{{\g}} \tilde{K} = \tr_{{\bg}}{\bar K}$ is the mean curvature, 
the physical extrinsic curvature (second fundamental form) $\bar{K}$ is related to  $ \tilde{K}$ (as contravariant tensors)  by
\[
\bar{K}=\varphi^{-\frac{2(n+2)}{n-2}} \tilde{K}+\frac{\tau}{n}\bar{\g}^{-1}
\]
with  $\bar{\g}^{-1}$ denoting the contravariant  form of the metric $\bar{\g}$, and where $\bar{J}:=-\pi \nabla \psi$. 
We have shown in \cite{CBIP}, following ideas originating from York \cite{Y99},
that it is useful to  associate to the  background conformal metric $\gamma$, a
function $\tilde{N}$ which is related to the original lapse by the
equation\footnote{This relation consists in requiring that each metric has
as associated to it an initial lapse with the same  ``densitized lapse''.}
\[
{N}(Det\bar{\g})^{-\frac{1}{2}}=\tilde{N}(Det\gamma)^{-\frac{1}{2}},
\]
or simply
\[
N=\vp^{\frac{2n}{n-2}} \tilde{N}.
\]
Then the ``physical'' scalar field initial data $(\bpsi, \bpi)$  consists of
$\bpsi = \psi$ 
and
\[
\bar{\pi}=r{N}^{-1}\overline{\partial}_{0}\psi=\varphi^{-\frac{2n}{n-2}}
{\pi}\quad\text{where}\qquad \pi=\tilde{N}^{-1}\partial_{0}\psi,
\]
and it follows that  
\[
\bar{J}=-\varphi^{-\frac{2(n+2)}{n-2}}\pi\nabla\psi=\varphi
^{-\frac{2(n+2)}{n-2}}J,\qquad  \text{where}\qquad J = -\pi\nabla\psi.
\]
Hence \textit{the Einstein-scalar field momentum constraint equation does not contain} 
$\varphi$ \textit{\ if } $\nabla\tau=0$.
Setting
\begin{equation}
\tilde{K}=\mathcal{L}_{\gamma, conf}X+U,\qquad\Delta_{\gamma, conf}:=\di_{\gamma}\mathcal{L}_{\gamma, conf}
\label{detK}
\end{equation}
with $\mathcal{L}_{\gamma, conf}X$ the conformal Lie derivative of $\gamma$ (or conformal Killing operator)
with respect to a vector field $X$ and with $U$ a freely specified traceless 2-tensor, the  system
may be regarded as a self-adjoint linear elliptic system for $X$, as follows.
\beq
\Delta_{\gamma, conf}X=-\di_{\gamma}U+\frac{n-1}{n}\varphi^{\frac{2n}{n-2}}
\nabla\tau-\pi\nabla\psi.
\label{confm}
\eeq
In summary, if we begin with a choice of  ``free" initial data $(\g, U, \tau, \psi, \pi)$ and solve the conformally 
formulated momentum constraint equation (\ref{confm}) to determine $\tilde{K}$ as indicated in (\ref{detK})
then $\tilde{K}$ satisfies the momentum constraint equation (\ref{momeqn}).

\subsubsection{The Hamiltonian constraint}  
If we specify the  initial data $(\g, U, \tau, \psi, \pi)$ and  use the identities  (\ref{cdiv}) and (\ref{cscal}),
then the  
Hamiltonian constraint equation (\ref{hce}) becomes a semilinear elliptic equation, called  the
Lichnerowicz equation, for $\varphi$. We have shown \cite{CBIP, CBIP2}  that it takes the form
\begin{equation}
\mathcal{H}\equiv\Delta_{\gamma}\varphi-f(\varphi)=0,
\label{lich}
\end{equation}
with 
\[
f(\varphi):= {\cal R}_{\gamma, \psi}\,\vp - {\cal A}_{\gamma,  \tilde{K}, \pi}\,\vp^{-\frac{3n-2}{n-2}}
+  {\cal B}_{\tau, \psi}\,\vp^{\frac{n+2}{n-2}}
\]
where we set $c_n := \frac{n-2}{4(n-1)}$,  and we let
\[
 {\cal R}_{\gamma, \psi}:=c_n\left(R(\gamma)-|\nabla{\psi}|^2_{\gamma}\right), \qquad
 {\cal A}_{\gamma, \tilde{K}, \pi}:=c_n\left(|\tilde{K}|^2_{\gamma}+\pi^2\right)
 \]
 and
 \[
{\cal B}_{\tau, \psi}:=c_n\left(\frac{n-1}{n} \tau^2 -4V(\psi) \right).
\]
We observe that ${\cal A}_{\gamma, \tilde{K}, \pi}\geq0,$ while the sign of ${\cal B}_{\tau, \psi}$ depends on the relative values of $\tau$ and $V(\psi)$.  Note that in the constant mean curvature case the system of equations are ``semi-decoupled" in 
that we may first solve the momentum constraint equation (\ref{momeqn}) (which does not involve $\vp$)
and then use the resulting $\tilde{K}$ to formulate the Lichnerowicz equation as described above.  If we can 
find a positive solution $\vp$ to this equation then this determines the physical metric $\bar{\g}$ and second fundamental
form $\bar{K}$ as well as the physical initial data $(\bar{\psi}, \bar{\pi})$ for the scalar field.

In \cite{CBIP, CBIP2} we establish a number of results regarding the existence or non existence of solutions  
for the system consisting of (\ref{confm}) and (\ref{lich}).  We present here some of the existence results, in 
a low regularity setting, on manifolds with $\tau$ constant (constant mean curvature, or  ``CMC", initial data).
The assertion that these results  hold in a low regularity setting follows from methods established 
by Choquet-Bruhat \cite{CB04}
and Maxwell \cite{Max2, Max3} for the vacuum Einstein constraint equations.

\subsection{Existence theorems.}

We denote by $W_{s}^{p}$ the usual Sobolev space on $(\Sigma, \gamma)$, consisting of functions with
(for $s$ a positive integer) all weak derivative of orders less than or equal to $s$ lying in $L^{p}$, and by
$M_{s}^{p}$ the space of $W_{s}^{p}$ Riemannian metrics (which consists of an open cone in
the space of all $W_{s}^{p}$ 2-tensors if $s\geq2$ and $p>\frac{n}{2}$, or if $p=2$ and $s>\frac{n}{2})$.
We denote $W_{s}^{2}$ by $H_{s}$ and  $M_{s}^{2}$ by $M_{s}$.

\subsubsection{Solving the momentum constraint.}

Given a traceless tensor $U$ on $\Sigma$, the conformally formulated momentum constraint 
equation (\ref{confm}) is a linear elliptic equation for the vector field $X$. 
The kernel of $\Delta_{\gamma, conf}$ consists of the space of conformal Killing vector fields.
The inhomogeneous term $\di_{\gamma}U$ is orthogonal to this space. 
The proof of the following theorem follows from known theorems for linear elliptic systems.
We suppose that $s$ is an integer.\footnote{The case of non
integer $s>\frac n2$ is treated by Maxwell \cite{Max3} for the vacuum Einstein constraint equations.
This of course requires working with distributional solutions if n=3. For integral choices of $s>\frac{n}{2}$ 
we in particular have $s\geq2$ when $n\geq3$. Thus the formulas presented here involve 
pointwise almost everywhere defined derivatives.}
We let $F$ denote the left hand side of (\ref{confm}) in the CMC setting, so that
\[
F=-\di_{\gamma}U-\pi\nabla\psi.
\]
\begin{theorem}
The conformally formulated momentum constraint equation  (\ref{confm}) with 
$\gamma\in M_{2}^{p}$, $p>\frac{n}{2}$ has a
solution $X\in W_{2}^{q}$, $1<q\leq p$, if $F\in L^{q}$ and if $J=\pi\nabla\psi$ is
orthogonal in the $L^{2}$ sense to the space of conformal Killing (CK)
vector fields on ($\Sigma, \gamma)$.

Moreover the solution is uniquely determined up to the addition of a conformal 
Killing vector field. There exists
 a constant $c(\g)$, depending only on $\gamma$, such that the unique 
solution which is orthogonal to the space of CK vectors
satisfies
\begin{equation}
\|X\|_{W_{2}^{q}}\leq c(\g)\| F\|_{L^{q}}.
\end{equation}
\end{theorem}

\begin{corollary}
If $\gamma\in M_{s}^{2},$ $s>\frac{n}{2}$ and $F\in H_{s-2},$ then $X\in H_{s}.$
\end{corollary}

\begin{proof}
We first remark that if $s>\frac{n}{2},$ then the Sobolev embedding theorem
$W_{2}^{p}\subset H_{s}$ for $p\leq \frac{2n}{n-2s+4}$
implies that if $s>\frac{n}{2}$ there exists $p>\frac{n}{2}$ such that the
embedding holds.

The corollary is then established in the usual way, by differentiating the equation and using the
Sobolev multiplication and interpolation properties.
\end{proof}

\subsubsection{Solving the Hamiltonian constraint.}

Satisfying the Hamiltonian constraint is equivalent to finding a positive
solution of the Lichnerowicz equation (\ref{lich}).
To prove the existence of positive solutions we use the method of sub and supersolutions.
The early approaches to solving the Lichnerowicz equation made use of the Leray-Schauder degree.
The method employed here uses estimates for linear elliptic equations together with the
Arzela-Ascoli theorem; 
see for example \cite{I95}. 
The first solutions were  found in
H\"older spaces, then in Sobolev spaces $H_{s}$, $s>\frac{n+1}{2}+1$. The
regularity has since been reduced to $W_{2}^{p},$ $p>\frac{n}{2}$ \cite{CB04} and to
$H_{s}$, $s>\frac{n}{2}$ \cite{Max3}, in the absence of scalar field.

We give a general theorem, adapted to equations of the type of the Lichnerowicz equation,
with or without a scalar field. Consider the semi-linear equation 
\begin{equation}
\Delta_{\gamma}\varphi=f(x,\varphi)\equiv\sum_{i=1,\ldots,N}a_{i}(x)\varphi
^{p_{i}},
\label{slpde}
\end{equation}
on the compact manifold $(\Sigma,\gamma)$,  where $x\in \Sigma$, and $p_{i}\in\R$.
We say that $\varphi_{-}$ is a subsolution of  (\ref{slpde})  if $\Delta_{\gamma}\varphi_-\geq f(x,\varphi_-)$ and 
$\varphi_{+}$ is a supersolution of  (\ref{slpde})  if $\Delta_{\gamma}\varphi_+\leq f(x,\varphi_+)$.
\begin{theorem}
Equation (\ref{slpde}) 
admits a positive solution $\varphi\in W_{2}^{p}$, $p>\frac{n}{2}$, provided the following
conditions are satisfied
\begin{itemize}
\item[(a)] $\gamma\in M_{2}^{p}$, $p>\frac{n}{2}$, and  $a_{i}\in L^{p}$ for  $i=1,\ldots,N$.
\item[(b)] The equation admits a strictly positive subsolution $\varphi_{-}$ and  supersolution
$\varphi_{+}$, both in $W_{2}^{p}$, with $0<\varphi_{-}\leq\varphi_{+}<\infty$. 
\end{itemize}
The solution $\varphi$ then  satisfies $\varphi_{-}\leq\varphi\leq\varphi_{+}$, and 
is unique if $f(x,y)$ is monotonically increasing in $y$ for each $x\in\Sigma$.
On the other hand, if all of the $a_{i}$ are of the same sign then no positive solution exists.
\end{theorem}

\begin{corollary}
If in addition $\gamma\in M_{s},$ $s>\frac{n}{2},$ a$_{i}\in H_{s-2},$
i=1,...,N, then $\varphi\in H_{s}.$
\end{corollary}
It can be proved by using conformal invariance that, in the case of the
Lichnerowicz equation, the uniqueness of the solution is independent of the
sign of $\lr$ (see Theorem 7.12 of \cite{CB06}).

In the original analysis of the Lichnerowicz equation for vacuum CMC data 
on compact manifolds, the full Yamabe theorem\footnote{This says that every metric
on a compact manifold is conformal to one with constant scalar curvature. The proof
of this  theorem  was completed by Schoen \cite{S84} after essential contributions by
Yamabe, Trudinger and Aubin.  We refer the interested reader to \cite{LP, CBDM} and the references
contained therein.} is employed
to fix the sign of the linear zero order term \cite{I95}.  A verification that one only needs
to control the {\em sign} of the scalar curvature (a much easier result), even in the low regularity setting, 
is provided by results of \cite{CB04} and \cite{Max3}.   The following result provides the analog of this 
control in the presence of a scalar field  (see Proposition 1 of \cite{CBIP2}).

\begin{theorem} [The Yamabe-scalar field conformal invariant]
 The functional on $H_{1}(\Sigma)$ (for $\gamma\in M_{2}^{p}$ and $\psi\in W_{2}^{p}$ given)
 defined by 
\begin{equation}
Q_{\gamma, \psi}(u) =\frac{{\displaystyle c_n^{-1} \int_\Sigma [|\nabla u|^2_\gamma + 
\lr u^2] \,d{\rm vol}_{\gamma}}}{{\displaystyle
\left( \int_\Sigma u^{\frac{2n}{n-2}}\,d{\rm vol}_{\gamma}\right)^{\frac{n-2}{n}} }}
\end{equation}
admits an infimum, $\yam>-\infty$, which is a conformal invariant. Its sign determines the
Yamabe-scalar field classes of pairs $(\gamma, \psi)$. 
A pair $(\gamma, \psi)$ with $\yam<0$ (respectively $\yam=0$, or $\yam>0$)
can be conformally transformed to a pair such that $\lr<0$ (respectively $\lr=0$,
or $\lr>0$) on $\Sigma$, and moreover if $\lr$ maintains a fixed sign on $\Sigma$ it is
necessarily of the same sign at $\yam$.
\label{classes}
\end{theorem}

The proof of the following existence theorem relies on the construction of
sub and supersolutions $\varphi_{-}$ and $\varphi_{+}.$ 
We have supposed that $\tau$ is a constant and that $V$ is a smooth function
of $\psi\in W_{2}^{p}\subset C^{0}(\Sigma)$, since $p>\frac{n}{2}$.  We therefore  also 
have $\lb\in C^{0}(\Sigma) \subset L^{\infty}$.

The expression  ${\cal A}_{\gamma, \tilde{K}, \pi}=c_n\left(|\tilde{K}|^2_{\gamma}+\pi^2\right)$
shows that $\la\geq0$, and that, for a solution $\tilde{K}$ of the momentum
constraint, $\la\in L^{p}$ since
\[
W_{1}^{p}\times W_{1}^{p}\subset L^{p}\text{ \ \ when \ \ }p>\frac{n}{2}.
\]
We present  results here in  the case that  
${\cal B}_{\tau, \psi}=c_n\left(\frac{n-1}{n} \tau^2 -4V(\psi) \right)\geq0$.
We refer the interested reader to \cite{CBIP2} for a more general treatment.
\begin{theorem}
Suppose that $\gamma\in M_{2}^{p}$, $p>\frac{n}{2}$ and that $\psi\in W_{2}^{p}$, 
$\tilde{K}, \pi \in W_{1}^{p}$ and $\lb\geq0$. Then the Lichnerowicz
equation (\ref{lich}) admits a positive solution $\varphi>0$, $\varphi\in W_{2}^{p}$, in the
following cases.
\begin{itemize}
\item[(1)] $(\gamma, \psi)$ is in the positive Yamabe-scalar field  class and $\la\not \equiv0$, or
\item[(2)] $(\gamma, \psi)$ is in the zero Yamabe-scalar field class and  $\underset{\Sigma}\inf\,\lb>0$.
\end{itemize}
\end{theorem}

\begin{proof}
If $(\gamma,\psi)$ is in the positive or zero Yamabe-scalar field  class a constant
supersolution can be constructed directly as follows.   First note that, as indicated in 
Theorem \ref{classes}, we may assume that $\lr\geq 0$.  This may require that we make a
preliminary conformal transformation of our initial data, and find a new solution of the conformally
formulated momentum constraint equation (see \cite{CBIP2}).
We consider the function of 
the single variable $y$ defined by 
\begin{equation}
F(y)={\ulb}y^{\frac{n}{n-2}}+{\ulr}y^{\frac{n-1}{n-2}}-{\ula},
\end{equation}
where \underline{$f$} denotes the mean value of a function $f$ on
$(\Sigma,\gamma):$
\[
\underline{f}\equiv\frac{1}{Vol(\Sigma,\gamma)}\int_{\Sigma}f d{\rm vol}_{\gamma}.
\]
Note that by setting $y(x)=\varphi(x)^4$ we see that $f(x,\phi)=y^{-\frac{3n-2}{4(n-2)}}F(y)$, provided that we 
do not replace the coefficients by their average values.  

Under the stated hypothesis one easily sees that 
$F(y)$ is monotonically increasing on $\R_+$ and has exactly one positive root.   
We let $y_{0}=\varphi_0^4$ denote this root, so that $F(y_0)=0$.
Now consider the linear equation
\begin{equation}
\Delta_{\gamma}v=\lr\varphi_{0}-\la\varphi_{0}^{-\frac{3n-2}{n-2}}+\lb\varphi_{0}^{\frac{n+2}{n-2}}.
\label{lineq}
\end{equation}
By our choice of $\varphi_0$ the right hand side of this equation has mean value zero and is therefore
orthogonal to the constants.  Thus we may consider the function $v\in W_{2}^{p}$, with mean value zero on $\Sigma$,
which solves (\ref{lineq}).
The function
\[
\varphi_{+}\equiv\varphi_{0}+v-\inf_{\Sigma}v\geq\varphi_{0},\qquad \Delta_{\gamma
}\varphi_{+}\equiv\Delta_{\gamma}v,
\]
is a supersolution if $\lr\geq0$ because it holds that:
\[
\Delta_{\gamma}\varphi_{+}-f(\cdot,\varphi_{+})=\lr(\varphi_{0}-\varphi_{+})-
\la(\varphi_{0}^{-\frac{3n-2}{n-2}}-\varphi_{+}^{-\frac{3n-2}{n-2}})+\lb(\varphi_{0}^{\frac{n+2}{n-2}}-\varphi_{+}^{\frac{n+2}{n-2}}).
\]
Hence if $\lr\geq0$, then since $\la\geq 0$ and $\lb\geq 0$, we have  
\[
\Delta_{\gamma}\varphi_{+}-f(\cdot,\varphi_{+})\leq0
\]
because $\varphi_{+}\geq\varphi_{0}$.
In the case of the zero  Yamabe-scalar field  class the same type of argument holds,
but we must use, in addition, the hypothesis $\lb\not \equiv0$ to insure that
$\varphi_{0}>0$.

In order to find a positive subsolution first note that 
any number $\ell<1$ such that
\[
\ell<\frac{\inf_{\Sigma}\la}{\sup_{\Sigma}(\lr+\lb)}
\]
is a constant subsolution. It is positive only if $\inf_{\Sigma}\la>0$. 

One may relax this
hypothesis on $\la$ by instead constructing a non constant subsolution using the
conformal invariance of the Lichnerowicz equation \cite{CBY, I95, Max3, CBIP2}.
We ignore for the purposes of this theorem the connection between the coefficient
$\la$ and the tensor $\tilde{K}$ arising from  the solution
to the conformally formulated momentum constraint equation
(\ref{confm}).  In order to state the conformal invariance properly  
one must, in addition to conformally rescaling the coefficients, solve (\ref{confm})
with the conformally transformed data before posing the conformally transformed Lichnerowicz
equation.  We refer the reader to Proposition 2 of \cite{CBIP2} for details.  Since we only 
are concerned here with the Lichnerowicz equation we may state the conformal invariance at follows:
\begin{eqnarray*}
&&\Delta_{\gamma}\varphi-\lr\varphi+\la\varphi^{-\frac{3n-2}{n-2}}-\lb\varphi
^{\frac{n+2}{n-2}}=\\
&&
\theta^{\frac{n+2}{n-2}}\left(\Delta_{\gamma^{\prime}}\varphi^{\prime}-\lr^{\prime}\varphi^{\prime}
+\la^{\prime}\varphi^{\prime-\frac{3n-2}{n-2}}-\lb^{\prime}\varphi^{\prime\frac{n+2}{n-2}}\right),
\end{eqnarray*}
with
\[
\gamma^{\prime}=\theta^{\frac{4}{n-2}}\gamma,\quad \varphi^{\prime}=\theta
^{-1}\varphi,\quad \la^{\prime}=\la\theta^{-\frac{4}{n-2}}\quad{\mbox{and}}\quad \lb^{\prime}=\lb.
\]
Now suppose $\la\geq0$, $\la\not \equiv0$. We set
\[
k=\lr+\lambda \lb
\]
with $\lambda=0$ in the positive Yamabe-scalar field case, while we take
$\lambda>(\inf_{\Sigma}\lb)^{-1}$ in the case $\lr=0$. Then there exists a $\theta>0$,
$\theta\in W_{2}^{p}$, such that
\[
\Delta_{\gamma}\theta-k\theta=-\la.
\]
Then $\Delta_{\gamma}\theta-\lr\theta=-\theta^{\frac{n+2}{n-2}}\lr^{\prime}$
implies
\[
\lr^{\prime}=\theta^{-\frac{n+2}{n-2}}\left(\la+(\lr-k)\theta\right).
\]
The ``primed Lichnerowicz equation" then admits the positive constant subsolution
$\ell$ if
\[
-\lr^{\prime}\ell+\la^{\prime}\ell^{-\frac{3n-2}{n-2}}-\lb^{\prime}\ell^{\frac{n+2}{n-2}}\geq0,
\]
or, equivalently,
\[
\theta^{-\frac{n+2}{n-2}}\{-(\la+\lambda \lb)\theta\}\ell+\la\theta^{-\frac{4}{n-2}}
\ell^{-\frac{3n-2}{n-2}}-\lb\ell^{\frac{n+2}{n-2}}\geq0.
\]
Any number $\ell$ such that
\[
\ell\leq \min(\inf_{\Sigma}\lambda^{\frac{n-2}{4}}\theta^{-1},\ \inf_{\Sigma}\theta^{\frac{n-2}{4(n-1)}}),
\]
is a positive subsolution of this transformed equation, and this shows that $\theta^{-1}\ell$ is a
positive subsolution of the original equation.
\end{proof}

An analogous method allows for the construction of sub and supersolutions, in the
negative Yamabe-scalar field class. We refer the interested reader to \cite{CB04} and
\cite{Max3} for the vacuum case and 
\cite{CBIP} for the Einstein-scalar field system.

\begin{remark}
The existence and non existence results presented here cover all cases when $\lb\geq0$ is a
constant,\footnote{This is the case when the
scalar field reduces to a cosmological constant.} since then either
$\inf_{\Sigma}\lb>0$ or $\lb\equiv0$, and one is therefore studying the equation
\[
\Delta\varphi\geq0\text{ \ \ \ [or }\leq0\text{] \ \ \ and }\not \equiv0,
\]
which has no solution on a compact manifold.
\end{remark}

\vfill\eject

\end{document}